\documentclass[sigconf,screen,authorversion,nonacm]{acmart}

\definecolor{classbg}{RGB}{220,235,255}
\definecolor{fieldbg}{RGB}{255,250,205}
\definecolor{methodbg}{RGB}{255,220,220}

\definecolor{linenumber}{gray}{0.4}

\AtBeginDocument{%
  }

\copyrightyear{2025} 
\acmYear{2025} 
\setcopyright{cc}
\setcctype{by}
\acmConference[SCORED '25]{Proceedings of the 2025 Workshop on Software Supply Chain Offensive Research and Ecosystem Defenses}{October 13--17, 2025}{Taipei, Taiwan}
\acmBooktitle{Proceedings of the 2025 Workshop on Software Supply Chain Offensive Research and Ecosystem Defenses (SCORED '25), October 13--17, 2025, Taipei, Taiwan}
\acmDOI{10.1145/3733827.3765530}
\acmISBN{979-8-4007-1915-8/2025/10}

\usepackage{xspace}
\usepackage{listings}
\usepackage{subcaption}
\usepackage{url}
\usepackage{multirow}

\newcommand{\deblometer}{\textsc{Deblometer}\xspace}

\definecolor[named]{OliveGreen}{rgb}{0,0.6,0}
\definecolor[named]{MidnightBlue}{cmyk}{1,0.1,0,0.1}
\definecolor[named]{Maroon}{rgb}{0.5, 0.0, 0.0}

\colorlet{punct}{red!60!black}
\definecolor{background}{HTML}{EEEEEE}
\definecolor{delim}{RGB}{20,105,176}
\colorlet{numb}{magenta!60!black}

\lstdefinelanguage{json}{
    basicstyle=\ttfamily\footnotesize,
    numbers=left,
    stepnumber=1,
    numbersep=8pt,
    showstringspaces=false,
    xleftmargin=5.5ex,
    breaklines=true,
    frame=tlrb,
    literate=
     *{0}{{{\color{numb}0}}}{1}
      {1}{{{\color{numb}1}}}{1}
      {2}{{{\color{numb}2}}}{1}
      {3}{{{\color{numb}3}}}{1}
      {4}{{{\color{numb}4}}}{1}
      {5}{{{\color{numb}5}}}{1}
      {6}{{{\color{numb}6}}}{1}
      {7}{{{\color{numb}7}}}{1}
      {8}{{{\color{numb}8}}}{1}
      {9}{{{\color{numb}9}}}{1}
      {:}{{{\color{punct}{:}}}}{1}
      {,}{{{\color{punct}{,}}}}{1}
      {\{}{{{\color{delim}{\{}}}}{1}
      {\}}{{{\color{delim}{\}}}}}{1}
      {[}{{{\color{delim}{[}}}}{1}
      {]}{{{\color{delim}{]}}}}{1},
}

\def\arraystretch{1.2}

\newif\ifshowcomments

\begin{document}

\title{A Soundness and Precision Benchmark for Java Debloating Tools}

\author{Jonas Klauke}
\affiliation{%
	\institution{Paderborn University}
	\city{Paderborn}
	\country{Germany}
}
\email{jonas.klauke@upb.de}
\orcid{0000-0001-9160-9636}

\author{Tom Ohlmer}
\affiliation{%
	\institution{Paderborn University}
	\city{Paderborn}
	\country{Germany}
}
\email{tohlmer@mail.upb.de}
\orcid{0009-0005-0644-7895}

\author{Stefan Schott}
\affiliation{%
	\institution{Paderborn University}
	\city{Paderborn}
	\country{Germany}
}
\email{stefan.schott@upb.de}
\orcid{0000-0002-0644-3297}

\author{Serena Elisa Ponta}
\affiliation{%
	\institution{SAP Labs}
	\city{Mougins}
	\country{France}
}
\email{serena.ponta@sap.com}
\orcid{0000-0002-6208-4743}

\author{Wolfram Fischer}
\affiliation{%
	\institution{SAP SE}
	\city{Stuttgart}
	\country{Germany}
}
\email{wolfram.fischer@sap.com}
\orcid{0000-0001-8127-8837}

\author{Eric Bodden}
\affiliation{%
	\institution{Paderborn University \& \\ Fraunhofer IEM}
	\city{Paderborn}
	\country{Germany}
}
\email{eric.bodden@upb.de}
\orcid{0000-0003-3470-3647}

\renewcommand{\shortauthors}{Klauke et al.}

\begin{abstract}
Modern software development reuses code by importing libraries as dependencies.
Software projects typically include an average of 36 dependencies, with 80\% being transitive, meaning they are dependencies of dependencies.
Recent research indicates that only 24.9\% of these dependencies are required at runtime, and even within those, many program constructs remain unused, adding unnecessary code to the project.
This has led to the development of debloating tools that remove unnecessary dependencies and program constructs while balancing precision by eliminating unused constructs and soundness by preserving all required constructs.
To systematically evaluate this trade-off, we developed \deblometer, a micro-benchmark consisting of 59 test cases designed to assess support for various Java language features in debloating tools.
Each test case includes a manually curated ground truth specifying necessary and bloated classes, methods, and fields, enabling precise measurement of soundness and precision.
Using \deblometer, we evaluated three popular Java debloating tools: Deptrim, JShrink, and ProGuard.
Our evaluation reveals that all tools remove required program constructs, which results in changed semantics or execution crashes.
In particular, the dynamic class loading feature introduces unsoundness in all evaluated tools.
Our comparison shows that Deptrim retains more bloated constructs, while ProGuard removes more required constructs.
JShrink’s soundness is significantly affected by limited support for annotations, which leads to corrupted debloated artifacts.
These soundness issues highlight the need to improve debloating tools to ensure stable and reliable debloated software.

\end{abstract}




\maketitle

\section{Introduction}
The software supply chain consists of all processes and components involved in the development, building, and delivery of software.
One part of it is the management of third-party libraries.
This is typically handled by package managers like Maven, npm, or PIP, which automatically download and integrate the dependencies of a modern software project.
These dependencies are either directly added by the developers, inherited from a parent project, or transitively added by other dependencies.
There are, on average, 36 dependencies in an industry software project, with $80\%$ being transitive~\cite{achilles}.  
However, Soto et al.~\cite{depclean} showed that $75.1\%$ ($2.7\%$ direct, $15.4\%$ inherited, $57\%$ transitive) are bloated and therefore, they are not executed in the project.
As a result, a large portion of dependencies in the software supply chain are maintained and updated without impacting the project itself.
Many researchers~\cite{StudyLessAttackSurface,attackSurfaceDeblo2, attackSurfaceDeblo,debloatingBenchJava,attackSurfaceDeblo3,attackSurfaceDeblo4} demonstrated that debloating successfully decreases the attack surface by removing vulnerabilities in the project dependencies.
However, if code required at runtime is incorrectly debloated (i.e., removed), it will cause execution failures~\cite{debloatingBenchmarkC,debloCrashApp}  or it may even lead to new vulnerabilities introduced by changed behavior during execution~\cite{StudyMoreAttack}.
This raises the demand for debloating tools that are \emph{sound}, ensuring all code required at runtime is preserved, and \emph{precise}, effectively removing all bloated code.

Debloating can be performed at different granularity levels.
One broader granularity is the complete removal of entire dependencies from the project; however, a more fine-grained form of debloating is the disposal of program constructs inside the dependencies.
Typical program constructs of Java dependencies are classes, methods, or fields.
These are also considered bloated if they are not accessed during the execution of a given software project.
For example, this could be caused by a dependency that provides capabilities to manage the file system, while the project only utilizes it to read files.
In this case, all classes, methods, and fields related to writing files become bloated program constructs from the perspective of the project.
Ponta et al.~\cite{projectKB} have mapped vulnerabilities to fix commits of open-source dependencies to pinpoint them to the actual program constructs, revealing that the vulnerability does not affect the entire dependency.
By removing these program constructs, the attack surface of the project is further reduced.
Bloated program constructs also pose a threat even if they are not executed, since they can be exploited by deserialization attacks~\cite{deserialAttack} that misuse the mechanism to execute them.

In this paper, we investigate the soundness and precision of debloating tools that are capable of removing bloated program constructs in Java.
We have created a manually curated micro-benchmark called \deblometer where we exactly know which program constructs are required and which are bloated. 
Similar to the micro-benchmark for call graphs developed by Reif et al~\cite{cats}, we evaluate the soundness and precision in \deblometer on encapsulated Java language features to directly spot the shortcomings of the debloating tools.
To find Java language features of interest, we have investigated current literature about debloating, established micro-benchmarks in similar problem areas and the Java language feature documentation resulting in 59 test cases highlighting the usage of 13 Java language features.
Each test case is evaluated on the correct removal of the following program constructs: classes, interfaces, methods, and fields.
The set of needed and unneeded program constructs is defined by a manually created ground truth for each evaluated language feature.
We evaluated Deptrim~\cite{deptrim}, JShrink~\cite{jshrink}, and ProGuard~\cite{proguard} on \deblometer. 
All tools support the removal of the evaluated program constructs. 

Our evaluation suggests that Deptrim tends to remove fewer program constructs, which results in a higher soundness score but a lower precision score. In contrast, ProGuard favors more aggressive debloating, resulting in a lower soundness score but a higher precision score.
The investigation of JShrink also reveals soundness and precision issues, especially due to its lack of support for Java annotations, which causes JShrink to produce corrupted JAR files that can no longer be executed.
All tools struggle with sound resolution of the dynamic class loading language feature, while JShrink and ProGuard also show no support for reflection language features.
The execution of unsound debloated test cases revealed crashes and altered behavior during execution.
This highlights the need for improvement in debloating tools to produce stable debloated dependencies without causing crashes or modifying program semantics.

This paper consists of the following contributions:
\begin{itemize}
	\item A micro-benchmark~\footnote{\url{https://github.com/secure-software-engineering/Deblometer}} for Java debloating tools that remove bloated classes, methods, and fields.  
	It consists of 59 test cases with manually curated ground truths categorized by 13 different language features.  
	\item A systematic evaluation~\footnote{\url{https://zenodo.org/records/16994277}} of the soundness and precision of three popular Java debloating tools, Deptrim, JShrink, and ProGuard, focused on the resolving of encapsulated Java language features.
\end{itemize}

\section{Approach}\label{sec:approach}
\begin{figure}[h]
    \centering
    \setlength{\fboxsep}{1px}
    \setlength{\fboxrule}{1px}
    \includegraphics[width=\linewidth]{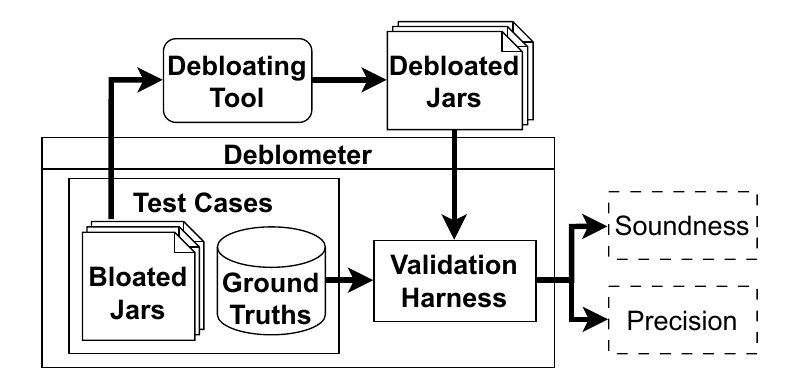}
    \caption{Overview of \deblometer}
    \label{fig:deblometer}
    \Description[Overview of the \deblometer, showing the input and the output of \deblometer and the internal structure]{Overview of \deblometer, showing the input and the output of \deblometer and the internal structure. The input is a debloated JAR and the output is soundness and precision scores. \deblometer consists of test cases with a corresponding ground truth which are compared in the validation harness of \deblometer.}
\end{figure}

An overview of \deblometer is shown in Figure \ref{fig:deblometer}.
It consists of the test cases and a validation harness.
Each test case is defined by a bloated JAR and its corresponding ground truth.
The debloating tool takes the bloated JARs as input, processes them, and produces the debloated Jars as output.
\deblometer evaluates these debloated JARs against their corresponding ground truths in the test cases within the validation harness.
The output of \deblometer includes the soundness and precision scores of the debloating tool for each test case.

% Ordner voller Jars debloated durch Deloater, enden in debloated Jars.

\subsection{Test Case Creation}
\label{subsec:testcreation}
The test cases assess the soundness and precision of the evaluated debloating tools in removing bloated program constructs.
Given the scope of our micro-benchmark, we categorized the test cases by language features to investigate the level of language feature support provided by the debloating tool.
Each test case involves a usage of one specific language feature.
It is specifically designed to use only that feature, allowing the soundness and precision results to be directly mapped to the corresponding language feature.
However, because some feature usages depend on the presence of another feature, not all test cases exclusively use a single language feature.
For example, Java supports functional interfaces, which combine lambdas and interfaces.
To evaluate this usage, we designed a test case categorized under interfaces, but it is also influenced by the lambda feature.
Since the test cases are used to evaluate debloating tools, they include additional classes, methods, and fields that are not involved in the usage of the targeted feature.
To construct a representative list of relevant language features for categorizing our test cases, we reviewed existing literature on debloating~\cite{jshrink,debloatingBenchJava} and static analysis~\cite{callgraphCompStudy,judge,cats,callSoundBench}, as many debloating tools rely on static analysis to process JAR files.
We also examined the Java language feature documentation~\cite{featureDoc} to identify features not discussed in the current literature.
A detailed overview of the selected language features is provided in Section \ref{sec:tests}.
For each language feature, we created a separate JAR file containing all corresponding test cases, including both required and bloated classes, methods, and fields.
These components are connected within the JAR by a main class that invokes all test cases.
This main class serves as the entry point for the debloating process.
It ensures that the debloating tools do not remove all program constructs in the bloated JAR simply because they are not reachable from the entry point.

\subsection{Debloating Levels}
\label{subsec:deblevel}
\deblometer consists of three debloating levels: class, method, and field.
The class level also includes interfaces and annotation definitions, as these are compiled into class files that are part of the bloated JAR.
Every test case contains program constructs from these three debloating levels, which may or may not be used during execution.
However, for certain language features, one or more types of program constructs may be excluded if they are not relevant to the specific feature being tested.
For example, in the case of method overloading, where multiple methods in a class can share the same name as long as their parameter lists differ, the field level is not considered because it is not related to the feature.

Listing \ref{fig:anonclass} presents an example test case from \deblometer.
This test case is categorized under the abstract class language feature.
In this example, the feature is exercised through the use of an anonymous class.
The program constructs for each debloating level are indicated by comments in the code.
Examples of these constructs include the abstract class \texttt{Car} in Line 3 for the class level, the integer field \texttt{piston} in Line 4 for the field level, and the abstract method \texttt{engine()} in Line 6 for the method level.
Within the \texttt{main} method, an anonymous class is instantiated with two overriding methods.
Only the \texttt{engine} method is invoked in Line 21.
This method also accesses the \texttt{piston} field from Line 4.
As a result, the \texttt{material} method in Line 18 and the \texttt{material} field in Line 5 are considered bloated because they are never used during program execution and can be removed.

\begin{lstlisting}[caption={Anonymous class implementation within the abstract module.},label=fig:anonclass, escapechar=\%]
package Abstract;

abstract class Car { //class level 
  int piston = 4; //field level
  String material = "aluminium"; //field level  
  public abstract void engine(); //method level 
  public abstract void material(); //method level 
}

public class Main { //class level
  public static void main(String[] args) {
    Car obj = new Car() { //class level
      @Override 
      public void engine() { //method level 
        System.out.println("No. of pistons in engine " + piston); 
      }
      @Override
      public void material() { //method level 
        System.out.println("Engine's material " + material);
    }; 
    obj.engine(); 
  } 
}

\end{lstlisting}

\subsection{Ground Truth}
\label{subsec:groundtruth}
For the ground truth specification, a JSON format is used to ensure that the information is structured and accessible during the validation process.
The JSON structure is organized according to the three debloating levels defined in the previous section: class, method, and field.
Each level is further divided into "required" and "bloated" sections.

In the class section, each class is identified by its package and class name.
In the method section, each entry includes the identifier of the declaring class, the method name, the return type, and the parameter list.
In the field section, each entry specifies the declaring class and the field name.

The example shown in Listing \ref{fig:json_bloat} represents the ground truth for the code sample in Listing \ref{fig:anonclass}.
The class level begins at Line 2.
All classes in the test case are executed, so they are listed in the "required" section from Lines 4 to 6, while the "bloated" section in Line 8 remains empty.
The \texttt{Car} class is specified in Line 5 using the class name \texttt{Car} and the package name \texttt{Abstract}.
The method level begins at Line 10.
Only two methods are required in the test case, as only \texttt{main()} and \texttt{engine()} are invoked.
The \texttt{engine()} method is described in Line 12 with the class name \texttt{Main\$1}, the method name \texttt{engine}, the return type \texttt{void}, and an empty parameter list.
The other \texttt{engine()} method and the \texttt{material()} method are not executed, and are therefore considered bloated.
They are listed in the "bloated" section from Lines 16 to 18.
The field level begins at Line 21.
The test case includes one required field and one bloated field, since only one field is accessed during execution.
The required field is defined in Line 23 with the class name \texttt{Car} and the field name \texttt{piston}.
If any of the required program constructs are removed during debloating, the resulting application either crashes due to incomplete code or behaves incorrectly. 
For example, if a method that overrides a superclass method is removed, the execution changes because Java will call the method from the superclass instead. 
This happens because Java uses dynamic method dispatch, and without the overriding method, the method call is resolved to the one defined in the superclass.
\begin{lstlisting}[language=json, caption={JSON representation of required and bloated elements on class, method, and field level.},label=fig:json_bloat]
{
 "CLASS": {
  "required": [
    {"package": "Abstract", "name": "Main"},
    {"package": "Abstract", "name": "Car"},
    {"package": "Abstract", "name": "Main$1"}
   ],
  "bloated": []
 },
 "METHOD": {
  "required": [
    {"type": "Main", "name": "main", "return": "void", "param": "String[]"},
    {"type": "Main$1", "name": "engine", "return": "void", "param": ""}
   ],
  "bloated": [
    {"type": "Car", "name": "engine", "return": "void", "param": ""},
    {"type": "Car", "name": "material", "return": "void", "param": ""},
    {"type": "Main$1", "name": "material", "return": "void", "param": ""}
   ]
 },
 "FIELD": {
  "required": [
    {"class": "Car", "name": "piston"}
   ],
  "bloated": [
    {"class": "Car", "name": "material"}
   ]
 }
}
\end{lstlisting}

\subsection{Validation Harness}
\label{subsec:validation}
The validation process analyzes the debloated JAR files produced by each debloating tool when applied to the bloated JARs from the \deblometer benchmark.
The remaining classes, methods, and fields in the debloated JAR are compared to the ground truth of the corresponding test case.
\deblometer uses SootUp~\cite{SootUp}, a static analysis framework that provides functionality to list all classes along with their declared fields and methods, to extract these program constructs.
The extracted elements are then compared to the required and bloated entries specified in the ground truth.

\deblometer identifies true positives, false positives, and false negatives at each debloating level in order to compute the soundness and precision for the class, method, and field levels.
A true positive refers to a program construct that is retained in the debloated JAR and correctly listed in the required section of the ground truth.
A false positive is a bloated construct that remains in the debloated JAR, even though it should have been removed.
A false negative occurs when a required construct is missing from the debloated JAR, which can lead to execution failures.

Soundness is calculated using the following formula:
\[
\text{Soundness} = \frac{TP}{TP + FN}
\]
Precision is calculated as:
\[
\text{Precision} = \frac{TP}{TP + FP}
\]
The evaluation results for each Java language feature are presented in tabular format, showing the soundness and precision at each of the three debloating levels, grouped by the language feature evaluated by the validation harness.

\section{Tests} \label{sec:tests}
In the following, we present the Java language features evaluated in \deblometer.
The process used to identify these features is described in Section \ref{subsec:testcreation}.
For each selected feature, we created multiple test cases to capture different usages of the evaluated language feature.
These test cases include both required and intentionally bloated program constructs.
The expected outcome of each test case is formally specified in the corresponding ground truth JSON files, as explained in Section \ref{subsec:groundtruth}.
These ground truths serve as the basis for evaluating the soundness and precision of the debloating tools, as detailed in Section \ref{subsec:validation}.
Table \ref{tab:testcases_amount} shows the number of test cases associated with each evaluated language feature.
In total, \deblometer contains 59 test cases categorized across 13 different Java language features.
All test cases categorized under a specific language feature are combined into a single bloated JAR representing that feature.

\begin{table}[htbp]
	\centering
	\caption{Number of test cases per Java language feature. A language feature is only considered evaluated in related work if it was explicitly investigated in that work.}
	\label{tab:testcases_amount}
	\begin{tabular}{@{}lp{3cm}c@{}}
		\textbf{Language Feature}&\textbf{Evaluated in Related Work}  & \textbf{Test Cases}  \\ \midrule
		Abstract                 &  & 6 \\
		Annotation               & & 7 \\
		Deserialization          & \cite{judge,cats, callSoundBench}&2 \\
		Dynamic Class Loading    & \cite{judge,cats,callSoundBench}&2 \\
		Exception                & &4 \\
		Externalization          & \cite{judge,cats}&2 \\
		Generics                 & \cite{callgraphCompStudy}&7 \\
		Interface                & \cite{judge,cats}&4 \\
		Lambda                   & \cite{callgraphCompStudy,judge,cats,callSoundBench}&4 \\
		Overloading              & \cite{judge,cats}&6 \\
		Overriding               & \cite{callgraphCompStudy,judge,cats}&4 \\
		Reflection               & \cite{callgraphCompStudy,judge,cats,callSoundBench}&6 \\
		Serialization            & \cite{judge,cats,callSoundBench}&5 \\
		 \midrule
		 & & 59 \\
	\end{tabular}
\end{table}

\subsection{Abstract Classes}
Abstract classes are a fundamental feature of object-oriented programming in Java, serving as templates for subclasses and enabling polymorphism.
To evaluate this language feature, we designed test cases based on various implementation techniques, including the use of anonymous classes, non-abstract subclasses, explicit method implementations, and scenarios with no usage of the abstract class.
Additional test cases include partial implementations that combine abstract and concrete logic, as well as unrelated method invocations not defined in the abstract class.

\subsection{Annotations}
Annotations are used to provide metadata that can influence program behavior at compile time, runtime, or during tooling processes.
To evaluate this feature, we designed test cases based on various implementation techniques, including the conditional presence of annotations on classes, fields, and methods; the use of functional annotations that affect program execution; and the identification of redundant annotations that do not impact program semantics.
Custom annotations were employed for this purpose, as standard Java annotations, while accessible, do not meet the specific requirements needed to support the analysis and removal of bloated code by debloating tools.

\subsection{Deserialization}
Deserialization enables the reconstruction of objects from a byte stream and plays a crucial role in data exchange and persistence.
To evaluate this language feature, we designed test cases based on relevant implementation techniques, including standard deserialization using built-in mechanisms and method overriding during deserialization.
We did not specifically test class-level debloating for this feature, as standard class reachability already applies, making such targeted testing redundant and unlikely to yield additional insights.
However, classes involved in deserialization are still covered in our evaluation through general usage scenarios, even if not tested through dedicated implementations.

\subsection{Dynamic Class Loading}
Dynamic class loading allows classes to be loaded at runtime, enabling greater extensibility and modularity.
To evaluate this language feature, we created test cases based on relevant implementation techniques, including dynamic instantiation using a class loader and reflective instantiation based on a specified class name.

\subsection{Exception}
Exception handling is used to manage errors and unexpected conditions during program execution.
To evaluate this language feature, we developed test cases based on different implementation techniques, including a basic custom exception, an extended custom exception with additional fields, a hierarchical structure of custom exceptions, and an implementation of an unchecked exception.

\subsection{Externalization}
Externalization provides developers with full control over the serialization process by allowing them to explicitly define how objects are written to and read from a stream.
This requires implementing the \texttt{readExternal} and \texttt{writeExternal} methods from the Externalizable interface.
To evaluate this language feature, we created test cases based on specific implementation techniques: one focused solely on reading and another solely on writing.

\subsection{Generics}
Generics enable type-safe programming by allowing classes, interfaces, and methods to operate on typed parameters.
They enhance code reusability and reduce the need for explicit type casting.
To evaluate this language feature, we developed test cases based on various implementation techniques, including basic generic type definitions, generic inheritance, generic interfaces, interface composition involving generics, method overloading with generic parameters, the use of parameterized arrays, and wildcard types for flexible type bounds.

\subsection{Interface}
Interfaces are a fundamental construct used to define expected behaviors that classes can implement, promoting abstraction and multiple inheritance of type.
To evaluate this language feature, we created test cases based on various implementation techniques, including standard interface implementation, the use of anonymous classes, interface composition to combine multiple behaviors, and lambda expressions representing functional interface implementations.

\subsection{Lambda Functions}
Lambdas are a language feature that enable a concise representation of functional behavior, allowing functions to be treated as first-class citizens.
To evaluate this feature, we designed test cases based on relevant implementation techniques, including lambda expressions that interact with classes, modify fields within their scope, and implement logic directly within methods.

\subsection{Overloading}
Overloading allows multiple methods with the same name to coexist within a class, distinguished by their parameter lists.
To evaluate this language feature, we designed test cases covering various implementation techniques, including methods with different numbers of arguments, compile-time binding resolution, variations in parameter types, swapped parameter order, method selection priority, and type promotion.
Class-level debloating was not specifically tested for this feature, as standard class reachability applies, making targeted testing redundant.
However, classes are implicitly covered through general program behavior, even though specific class implementations were not individually evaluated.
Additionally, since overloading applies only to methods, testing fields was not necessary.

\subsection{Overriding}
Overriding allows a subclass to provide a specific implementation of a method defined in its superclass, enabling dynamic dispatch and runtime polymorphism.
To evaluate this language feature, we designed test cases based on relevant implementation techniques, including standard method overriding, polymorphic behavior through superclass references, and method hiding where static methods in a subclass obscure rather than override those in the superclass.
Field-level testing was not performed because overriding applies strictly to methods and does not affect fields.

\subsection{Reflection}
Reflection enables inspection and manipulation of classes, methods, and fields at runtime, allowing dynamic behavior and introspection.
To evaluate this language feature, we created test cases based on various implementation techniques, including access control checks using name equality, retrieval of enum-defined fields and methods, inheritance-aware introspection, member retrieval, method inspection via enums, and standard reflection involving iteration over fields and methods.

\subsection{Serialization}
Serialization enables objects to be converted into a byte stream for storage or transmission, preserving their state for later reconstruction through deserialization.
To evaluate this language feature, we designed test cases based on relevant implementation techniques, including deep serialization of classes containing other serializable objects, serialization through inheritance hierarchies, and standard serialization scenarios where methods or logic without state are considered obsolete, as they cannot be meaningfully serialized.

\section{Evaluation}
In this section, we evaluate the behavior and limitations of debloating tools, measuring their soundness and precision using \deblometer.
Because our focus is on Java and the removal of program constructs, we selected tools that can debloat Java applications by removing classes, methods, and fields.
To identify suitable tools, we surveyed available Java debloating solutions that meet these criteria and are publicly accessible.
\deblometer evaluates the following tools\footnote{We initially included J-Reduce~\cite{jreduce} in our evaluation. However, it consistently produced empty JAR files for our test cases, so we excluded it from the final analysis.}:
\begin{itemize}
	\item Deptrim (v0.1.2)
	\item JShrink\footnote{\url{https://github.com/UCLA-SEAL/JShrink}} (retrieved on June 2, 2025)
	\item ProGuard (v7.7)
\end{itemize}

The selected tools use either static analysis, dynamic analysis, or a combination of both to identify all required program constructs.
ProGuard relies exclusively on static analysis during the debloating process, whereas Deptrim and JShrink combine static and dynamic analysis to address the known limitations of static analysis reported in related work \cite{cats,judge,callSoundBench}.
However, the default configuration of JShrink, as suggested by its authors, relies solely on static analysis for debloating.

\begin{figure}[h]
	\centering
	\setlength{\fboxsep}{1px}
	\setlength{\fboxrule}{1px}
	{\includegraphics[width=\linewidth]{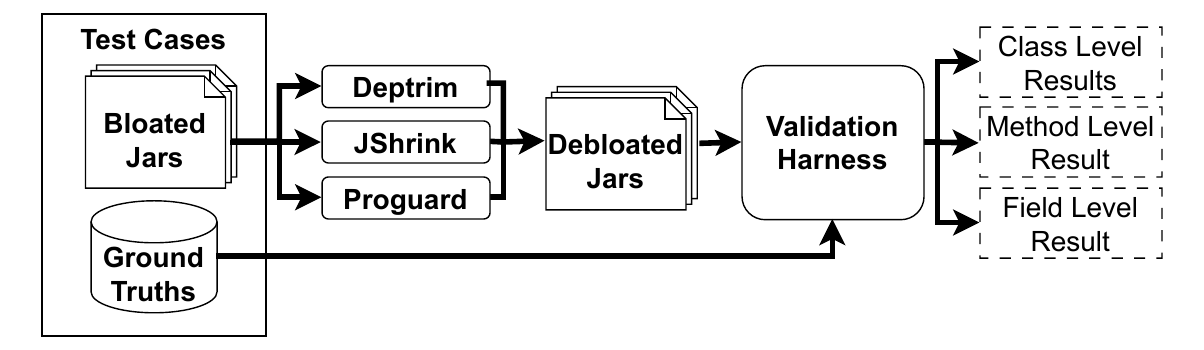}}
	\caption{The evaluation process using \deblometer}
	\label{fig:EvalProc}
	\Description[Overview of the evaluation process, showing the input and the output of \deblometer]{Overview of the evaluation process, showing the input and the output of \deblometer. The input is a bloated JAR which is debloated by Deptrim, JShrink, and Proguard. The debloated JAR is put into the validation harness of \deblometer which results in soundness and precision score for the class, method, and field level.}
\end{figure}

The evaluation process for the selected tools is illustrated in Figure \ref{fig:EvalProc}.
It begins by running the debloating tools on the bloated JARs provided by the test cases in \deblometer.
Each tool generates debloated JARs from the input JARs, which are then validated against the corresponding ground truths using \deblometer’s validation harness to calculate soundness and precision scores for each language feature and debloating level.
To ensure accurate validation, all obfuscation features were disabled, preserving original class, method, and field names.
Since Deptrim does not operate directly on JAR files but rather on project dependencies, we created a wrapper project that includes all \deblometer bloated JARs as dependencies, along with a main class that invokes the main methods of each dependency.
If Deptrim does not remove any program constructs and therefore produces no debloated JAR, we treat the original bloated JAR as its result.
When \deblometer identifies a language feature as unsoundly supported, we execute the corresponding debloated JAR to verify whether it remains functional.

Table~\ref{tab:benchmark_results} presents the results of our experiments.
It shows the soundness and precision results for each language feature, divided by the debloating levels of class, method, and field for the three debloating tools: Deptrim, JShrink, and ProGuard.
The table shows the grouped results of all test cases categorized under each specific language feature.
The column R lists the number of required classes, methods, or fields contained in the debloated JAR.
For example, 4/6 means that 4 of the required program constructs are present in the debloated JAR, while 2 are missing.
The column B lists the number of removed bloated classes, methods, or fields in the debloated JAR.
For example, 7/20 means that 7 of the bloated constructs were removed, while 13 remain in the debloated JAR.
The column S shows the soundness score as a percentage, and the column P shows the precision score as a percentage.
\begin{table*}[htbp]
	\centering
		\caption{Evaluation results at the class, method, and field levels for each language feature.
			The table shows the number of required program constructs (R), the soundness score (S), the number of removed bloated program constructs (B), and the precision score (P).
			S and P are shown as percentages and highlighted in bold if they are below 100.
			Method-level evaluation of serialization is omitted because there are no true positives in the ground truth.
			Similarly, overriding and overloading are not assessed at the field level, as the test cases do not include relevant fields.
			Entries marked with ‘–’ indicate a 0/0 result, except in explicitly excluded cases.}
	\resizebox{0.84\textwidth}{!}{%
		\small
		\setlength{\tabcolsep}{4pt}  % tighter columns if needed
		\renewcommand{\arraystretch}{1.1}	
	\begin{tabular}{ll|cccc|cccc|cccc}
		\multicolumn{2}{c|}{\textbf{Language Feature}} &
		\multicolumn{4}{c|}{\textbf{Deptrim}} &
		\multicolumn{4}{c|}{\textbf{JShrink}} &
		\multicolumn{4}{c}{\textbf{ProGuard}} \\[-0.2em]
		\multicolumn{2}{c|}{} &
		\textbf{R} & \textbf{S} & \textbf{B} & \textbf{P} &
		\textbf{R} & \textbf{S} & \textbf{B} & \textbf{P} &
		\textbf{R} & \textbf{S} & \textbf{B} & \textbf{P} \\
		\midrule
		\multirow{3}{*}{abstract} & Class  & 15/15 & 100 & 5/5 & 100 & 3/15 & \textbf{20} & 5/5 & 100 & 15/15 & 100 & 5/5 & 100 \\
		& Method & 5/5 & 100 & 7/20 & \textbf{48} & 1/5 & \textbf{20} & 19/20 & \textbf{50} & 5/5 & 100 & 19/20 & \textbf{83} \\
		& Field  & 2/2 & 100 & 2/3 & \textbf{80} & 1/2 & \textbf{50} & 2/3 & \textbf{50} & 2/2 & 100 & 2/3 & \textbf{67} \\
		\cmidrule{1-14}
		\multirow{3}{*}{annotation} & Class  & 24/24$_{b}$ & 100$_{b}$ & 0/4$_{b}$ & \textbf{86$_{b}$} & 4/24$_{c}$ & \textbf{17$_{c}$} & 4/4$_{c}$ & 100$_{c}$ & 24/24 & 100 & 4/4 & 100 \\
		& Method & 5/5$_{b}$ & 100$_{b}$ & 0/2$_{b}$ & \textbf{71$_{b}$} & 1/5$_{c}$ & \textbf{20$_{c}$} & 2/2$_{c}$ & 100$_{c}$ & 2/5 & \textbf{40} & 2/2 & 100 \\
		& Field  & 3/3$_{b}$ & 100$_{b}$ & 0/3$_{b}$ & \textbf{50$_{b}$} & 1/3$_{c}$ & \textbf{33$_{c}$} & 2/3$_{c}$ & \textbf{50$_{c}$} & 3/3 & 100 & 0/3 & \textbf{50} \\
		\cmidrule{1-14}
		\multirow{3}{*}{deserialization} & Class  & 4/4$_{b}$ & 100$_{b}$ & -$_{b}$ & 100$_{b}$ & 4/4 & 100 & - & 100 & 4/4 & 100 & - & 100 \\
		& Method & 3/3$_{b}$ & 100$_{b}$ & 0/2$_{b}$ & \textbf{60$_{b}$} & 3/3 & 100 & 0/2 & \textbf{60} & 1/3 & \textbf{33} & 2/2 & 100 \\
		& Field  & 4/4$_{b}$ & 100$_{b}$ & 0/1$_{b}$ & \textbf{80$_{b}$} & 4/4 & 100 & 0/1 & \textbf{80} & 4/4 & 100 & 0/1 & \textbf{80} \\
		\cmidrule{1-14}
		\multirow{3}{*}{dynamic class loading} & Class  & 4/6 & \textbf{67} & 4/4 & 100 & 4/6 & \textbf{67} & 4/4 & 100 & 4/6 & \textbf{67} & 4/4 & 100 \\
		& Method & 2/4 & \textbf{50} & 4/4 & 100 & 2/4 & \textbf{50} & 4/4 & 100 & 2/4 & \textbf{50} & 4/4 & 100 \\
		& Field  & 0/2 & \textbf{0} & 6/6 & \textbf{0} & 0/2 & \textbf{0} & 6/6 & \textbf{0} & 0/2 & \textbf{0} & 6/6 & \textbf{0} \\
		\cmidrule{1-14}
		\multirow{3}{*}{exception}  & Class  & 9/9 & 100 & 3/3 & 100 & 9/9 & 100 & 3/3 & 100 & 9/9 & 100 & 3/3 & 100 \\
		& Method & 1/1 & 100 & 0/1 & \textbf{50} & 1/1 & 100 & 1/1 & 100 & 1/1 & 100 & 1/1 & 100 \\
		& Field  & 1/1 & 100 & 0/1 & \textbf{50} & 1/1 & 100 & 1/1 & 100 & 1/1 & 100 & 1/1 & 100 \\
		\cmidrule{1-14}
		\multirow{3}{*}{externalization}  & Class  & 4/4 & 100 & 2/2 & 100 & 4/4 & 100 & 2/2 & 100 & 4/4 & 100 & 2/2 & 100 \\
		& Method & 2/2 & 100 & 4/6 & \textbf{50} & 2/2 & 100 & 4/6 & \textbf{50} & 2/2 & 100 & 4/6 & \textbf{50} \\
		& Field  & 4/4 & 100 & 4/6 & \textbf{67} & 4/4 & 100 & 6/6 & 100 & 4/4 & 100 & 6/6 & 100 \\
		\cmidrule{1-14}
		\multirow{3}{*}{generics}   & Class  & 20/20 & 100 & 3/4 & \textbf{95} & 20/20 & 100 & 4/4 & 100 & 18/20 & \textbf{90} & 4/4 & 100 \\
		& Method & 10/10 & 100 & 1/9 & \textbf{56} & 10/10 & 100 & 7/9 & \textbf{83} & 10/10 & 100 & 7/9 & \textbf{83} \\
		& Field  & 4/4 & 100 & 0/7 & \textbf{36} & 4/4 & 100 & 7/7 & 100 & 4/4 & 100 & 7/7 & 100 \\
		\cmidrule{1-14}
		\multirow{3}{*}{interface}  & Class  & 14/14 & 100 & 1/1 & 100 & 13/14$_{c}$ & \textbf{93$_{c}$} & 1/1$_{c}$ & 100$_{c}$ & 11/14 & \textbf{79} & 1/1 & 100 \\
		& Method & 5/5 & 100 & 1/7 & \textbf{45} & 5/5$_{c}$ & 100$_{c}$ & 6/7$_{c}$ & \textbf{83$_{c}$} & 5/5 & 100 & 6/7 & 83 \\
		& Field  & 1/1 & 100 & 0/1 & \textbf{50} & 0/1$_{c}$ & \textbf{0$_{c}$} & 1/1$_{c}$ & \textbf{0$_{c}$} & 0/1 & \textbf{0} & 1/1 & \textbf{0} \\
		\cmidrule{1-14}
		\multirow{3}{*}{lambda}     & Class  & 10/10 & 100 & 1/1 & 100 & 10/10$_{c}$ & 100$_{c}$ & 1/1$_{c}$ & 100$_{c}$ & 10/10 & 100 & 1/1 & 100 \\
		& Method & 4/4 & 100 & 1/2 & \textbf{80} & 4/4$_{c}$ & 100$_{c}$ & 2/2$_{c}$ & 100$_{c}$ & 4/4 & 100 & 2/2 & 100 \\
		& Field  & 2/2 & 100 & 0/2 & \textbf{50} & 2/2$_{c}$ & 100$_{c}$ & 0/2$_{c}$ & \textbf{50$_{c}$} & 2/2 & 100 & 0/2 & \textbf{50} \\
		\cmidrule{1-14}
		\multirow{3}{*}{overloading}& Class  & 14/14$_{b}$ & 100$_{b}$ & -$_{b}$ & 100$_{b}$ & 14/14 & 100 & - & 100 & 14/14 & 100 & - & 100 \\
		& Method & 8/8$_{b}$ & 100$_{b}$ & 0/7$_{b}$ & \textbf{53$_{b}$} & 8/8 & 100 & 7/7 & 100 & 7/8 & \textbf{88} & 7/7 & 100 \\
		& Field  & -$_{b}$ & -$_{b}$ & -$_{b}$ & -$_{b}$ & - & - & - & - & - & - & - & - \\
		\cmidrule{1-14}
		\multirow{3}{*}{overriding} & Class  & 11/11 & 100 & 1/1 & 100 & 11/11 & 100 & 1/1 & 100 & 11/11 & 100 & 1/1 & 100 \\
		& Method & 4/4 & 100 & 1/4 & \textbf{57} & 4/4 & 100 & 2/4 & \textbf{67} & 4/4 & 100 & 2/4 & \textbf{67} \\
		& Field  & - & - & - & - & - & - & - & - & - & - & - & - \\
		\cmidrule{1-14}
		\multirow{3}{*}{reflection} & Class  & 13/13$_{b}$ & 100$_{b}$ & -$_{b}$ & 100$_{b}$ & 13/13 & 100 & - & 100 & 13/13 & 100 & - & 100 \\
		& Method & 5/5$_{b}$ & 100$_{b}$ & 0/4$_{b}$ &\textbf{56$_{b}$} & 0/5 & \textbf{0} & 4/4 & \textbf{0} & 0/5 & \textbf{0} & 4/4 & \textbf{0} \\
		& Field  & 6/6$_{b}$ & 100$_{b}$ & 0/6$_{b}$ & \textbf{50$_{b}$} & 3/6 & \textbf{50} & 3/6 & \textbf{50} & 3/6 & \textbf{50} & 3/6 & \textbf{50} \\
		\cmidrule{1-14}
		\multirow{3}{*}{serialization} & Class  & 13/13 & 100 & 1/2 & \textbf{93} & 13/13 & 100 & 1/2 & \textbf{93} & 13/13 & 100 & 2/2 & 100 \\
		& Method & - & - & - & - & - & - & - & - & - & - & - & - \\
		& Field  & 14/14 & 100 & 2/4 & \textbf{88} & 14/14 & 100 & 2/4 & \textbf{88} & 14/14 & 100 & 4/4 & 100 \\
		\multicolumn{14}{l}{\textit{b:} bloated JAR was evaluated}\\
		\multicolumn{14}{l}{\textit{c:} \deblometer crashed due to corrupted bytecode; manual intervention}\\
	\end{tabular}%
	}
	\label{tab:benchmark_results}
\end{table*}

Regarding Table~\ref{tab:benchmark_results}, all three tools (Deptrim, JShrink, and ProGuard) demonstrate generally high soundness and precision at the class level.
Among them, Deptrim consistently yields the best overall results and performs especially well in cases involving annotations and generics, where JShrink and ProGuard show minor deficiencies.
In addition, JShrink produces corrupted JARs for the annotation, lambda, and interface language features, which causes \deblometer to crash because the class files cannot be parsed by SootUp.
A manual inspection revealed that the corruption was caused by an incomplete removal of annotations.
JShrink removed the annotation from the constant pool of the class file, but references to deleted constant pool values remained.
As a result, the corrupted JARs could not be executed or parsed by \deblometer.
To enable evaluation, we removed the corrupted class files from the affected debloated JARs.
The program constructs in these files were manually reviewed and their results were incorporated into the evaluation by \deblometer.
The missing support for annotations also affected the lambda and interface test cases.
This is because lambda classes include annotations in the class file, and one of the interface test cases makes use of lambdas as well.
It is important to note that not all language feature tests were specifically designed to evaluate class-level debloating.
In some cases, the tests rely only on basic class detection rather than class-specific usage of the language feature.
This can lead to inflated scores that do not fully reflect the actual debloating capabilities of the tools.
Despite these limitations, Deptrim proves to be the most robust and reliable tool for class-level analysis.
ProGuard performs well and ranks just behind, while JShrink falls short due to its incomplete annotation handling in certain test cases.

Regarding Table~\ref{tab:benchmark_results}, all three tools achieve generally high soundness scores at the method level.
This indicates that relevant methods are consistently preserved during the debloating process.
However, the precision scores across most language features reveal a significant shortcoming: the tools often retain a substantial amount of bloated code, leading to high false positive rates.
This shows that while the tools are effective at preserving required methods, they struggle to remove unnecessary ones with precision.
For instance, although Deptrim maintains perfect soundness across nearly all features, its precision remains relatively low in cases such as abstract classes (28 percent) and annotations (71 percent on bloated JARs), indicating excessive code retention.
In contrast, ProGuard applies a more precise debloating strategy and often outperforms Deptrim in terms of precision.
However, this increased precision comes at the cost of soundness in several cases.
This issue is especially evident in annotations, deserialization, and overloading, where required methods are incorrectly removed.
Such behavior reflects a more aggressive and risk-prone approach, prioritizing leaner outputs over complete correctness.
JShrink, by comparison, exhibits low or unstable precision across multiple features and occasionally crashes due to corrupted JARs.
Overall, the results highlight that while method-level soundness is generally satisfactory, achieving high precision without compromising correctness remains a key challenge.
Among the evaluated tools, ProGuard demonstrates the most aggressive trade-off, focusing on reducing code size even if it risks removing critical program elements.

Regarding Table~\ref{tab:benchmark_results}, the three debloating tools show consistently high soundness scores across most language features at the field level.
This indicates that they are generally effective at retaining required fields.
However, precision remains a persistent issue for all tools, with relatively high false positive rates suggesting difficulty in eliminating unused or irrelevant fields.
For example, although Deptrim achieves 100 percent soundness across nearly all features, its precision varies significantly.
It drops to 36 percent for generics and 50 percent for features such as annotations, interfaces, and lambdas.
In the interface feature, Deptrim is the only tool that maintains full soundness, while both JShrink and ProGuard fail entirely, with soundness and precision scores of 0 percent.
This result underscores Deptrim's strength in preserving essential field-level declarations.
Despite this, all tools fail in the dynamic class loading cases, with soundness and precision both at 0 percent.
Overall, these results show that identifying and retaining necessary fields is mostly successful, especially for Deptrim.
However, the accurate removal of bloated field-level code remains a common and unresolved challenge for all evaluated tools.

Finally, we executed all debloated JARs that \deblometer identified as unsound.
This included 1 JAR for Deptrim, 5 for JShrink, and 7 for ProGuard, each representing an unsoundly supported language feature.
The JARs produced by Deptrim and JShrink resulted in exceptions that caused the programs to crash.
Similarly, 3 of the 7 unsound JARs from ProGuard also crashed during execution.
The remaining 4 ProGuard JARs exhibited altered runtime behavior due to Java's dynamic resolution of virtual method calls.
Such behavioral changes are especially concerning, as they are difficult to detect and may introduce new vulnerabilities that do not exist in the original JAR.

\section{Discussion}
The main challenges that lead to unsoundness and imprecision in debloating arise from the limitations of static analysis when dealing with dynamic or implicit language features, as highlighted in related work on static analysis techniques~\cite{cats,judge,callSoundBench}.
Across all evaluation levels shown in the tables, one major source of unsoundness is the inability to correctly analyze reflective and dynamic behavior.
At the method level, reflection presents a particularly difficult challenge.
Only Deptrim achieves 100 percent soundness in these cases, thanks to its use of dynamic analysis that enhances static analysis by resolving dynamic features.
This approach demonstrates that achieving high soundness often requires sacrificing precision, and achieving high precision may come at the cost of soundness.
JShrink performs especially poorly when handling annotations.
It produces corrupted debloated JARs that crash during execution in several test cases.
In addition, it lacks support for Java annotations, which significantly limits its applicability for modern Java applications.
The challenges of static analysis are further demonstrated in the dynamic class loading test cases.
Here, required classes that are loaded at runtime by the classloader are incorrectly removed.
This failure not only breaks class-level soundness but also obscures actual field usage, making it harder to preserve essential fields.
As a result, tools such as JShrink and ProGuard fail to retain required fields, leading to broken runtime behavior and reduced soundness at both the method and field levels.
These examples show that current debloating approaches tend to either retain too much code, resulting in low precision, or remove too aggressively, leading to unsoundness.
This trade-off becomes particularly problematic when handling dynamic language constructs, reflection, or incomplete modeling of dependencies.

\section{Threats to Validity}
A potential threat to validity arises from possible inaccuracies in the manually created ground truth.
To minimize this risk, three authors independently verified the ground truth to ensure its accuracy.
Another threat is that the test cases are artificially constructed and may not reflect real-world scenarios.
However, the test cases in \deblometer are specifically designed to clearly expose shortcomings related to individual language features.
A further limitation is that the test cases might be too narrow to allow for generalization of the results.
To address this, we designed multiple test cases for each language feature and applied them in different contexts.
Another threat is that the set of language features evaluated in \deblometer does not cover every possible feature in Java.
To mitigate this, we focused on language features known to introduce unsoundness and imprecision, as reported in related work.
Another potential threat to validity lies in how the debloating tools were configured.
To reduce the impact of suboptimal configurations, we contacted the developers of each tool to ensure the most appropriate settings were used for our micro-benchmark scenarios.

\section{Related Work}
Most debloating tools are evaluated using popular benchmarks like GNU Coreutils~\cite{coreutils}, DaCapo~\cite{dacapo}, and SPEC CPU 2006/2017~\cite{spec06,spec17}, however, these benchmarks are not specially tailored for debloating approaches.
Chisel~\cite{chisel} and OCCAM~\cite{occam,occam2} have published benchmarks~\cite{chiselBench} for their tools in which real-world applications are debloated as test cases. 
Compared to \deblometer, these benchmarks spot either failures or investigate the code reduction, however, they do not inform the developers about specific language features that cause issues in the debloating process. 
Brown et al.~\cite{debloatingBenchmarkC} extended the chisel benchmark with more complex real-world applications and evaluated a set of debloating tools on it.
Their test cases are written in the C programming language, making them not applicable to Java debloating tools. The industrial readiness of Java debloating tools is analyzed by Ponta et al.~\cite{debloatingBenchJava}, however, their scope is on the attack surface reduction of vulnerable
Java dependencies, lacking insights into the soundness and precision of these tools.
Alhanahnah et al.~\cite{debloatingLandscape} surveyed the landscape of debloating literature and their performed evaluations, however, they did not compare the tools or evaluate their soundness and precision.
Java language feature support within call graph construction is analyzed by Reif et al.~\cite{cats,judge} and Sui et al.~\cite{callSoundBench,recallStudy} using micro-benchmarks and dynamic analysis.
Their approaches however are not fully applicable for evaluating debloating tools due to their scope being limited to method level.

\section{Conclusion}
In this paper, we presented \deblometer, a novel micro-benchmark designed to evaluate the debloating of program constructs in bloated JARs.
It consists of 59 test cases, each focused on a specific Java language feature.
We used \deblometer to assess the soundness and precision of three debloating tools capable of removing classes, methods, and fields in Java: Deptrim, JShrink, and ProGuard.
Our evaluation reveals clear trade-offs between soundness and precision, driven by fundamentally different tool strategies.
Deptrim prioritizes soundness through conservative analysis and performs soundly on nearly all evaluated language features except for dynamic class loading.
However, this approach results in imprecision across all features.
ProGuard employs a more aggressive debloating strategy, achieving higher precision at the cost of soundness, particularly for complex features such as annotations and deserialization.
It is also the only tool that is unsound on test cases involving generics and method overloading.
JShrink exhibits limited support for abstract classes and Java annotations, attaining the lowest scores on both features.
It can also produce corrupted JARs that fail to execute when the bloated JAR contains annotations.
Results from \deblometer highlight that modern debloating tools struggle significantly with dynamic language features such as reflection and dynamic class loading.
Method-level reflection and field usage under dynamic class loading expose the current limitations of these tools in balancing soundness and precision.
\deblometer can guide the development and improvement of debloating tools by precisely identifying soundness and precision issues related to specific Java language features.
These issues can then be addressed systematically by developers

\begin{acks}
We thank Subramanya Padmaraja Setty for his help with the design of the benchmark and for assisting in the gathering of language features and test cases.
This work was partially supported by the German Research Foundation (DFG) within the Collaborative Research Centre ”On-The-Fly Computing“ (GZ: SFB 901/3) under the project number 160364472.
\end{acks}

\bibliographystyle{ACM-Reference-Format}
\bibliography{paper}

\end{document}